\documentclass[10pt]{article}

\usepackage{amsmath}
\usepackage{amssymb}
\usepackage{dcolumn}
\usepackage{bm}

\usepackage{amsmath}
\usepackage{amsfonts}   
\usepackage{graphics}  
\usepackage{psfrag} 
\usepackage{graphicx} 
\usepackage{epsfig}    
\usepackage{rotating}  
\usepackage{lineno}  
 \usepackage[latin1]{inputenc}
\usepackage[usenames,dvipsnames]{color}
 \usepackage{rotating}

\usepackage{graphicx}

\usepackage{color} 
\definecolor{darkgreen}{rgb}{0,0.5,0}
\definecolor{Grey}{rgb}{0.5,0.5,0.5}
\definecolor{DarkYellow}{rgb}{1,0.7,0}
\definecolor{Violet}{rgb}{0.6,0.0,0.7}
\definecolor{Brown}{rgb}{0.5,0.3,0}

\usepackage[labelfont=bf,labelsep=period,justification=raggedright]{caption}

\date{}

\begin{document}
\begin{center}
{
{\Large
~\vspace{0.5cm}~

\textbf{The Emergence of Cooperation from a Single Mutant during Microbial Life-Cycles}
}
\\~\\
Anna Melbinger$^{1,2,\ast}$, Jonas Cremer$^{1,2}$, Erwin Frey$^{1}$
}\\~\\
{\bf 1} Arnold Sommerfeld Center for Theoretical Physics and Center for NanoScience, Department of Physics, Ludwig-Maximilians-Universit\"at M\"unchen, Germany
\\
{\bf 2} Department of Physics, UCSD, 9500 Gilman Drive, La Jolla, CA, USA
\\
$\ast$  E-mail: amelbinger@ucsd.edu
\end{center}
\vspace{1cm}
\begin{abstract}
Cooperative behavior is widespread in nature, even though cooperating individuals always run the risk to be exploited by free-riders.
Population structure effectively promotes cooperation given that a threshold in the level of cooperation was already reached. However, the question how cooperation can emerge from a single mutant, which cannot rely on a benefit provided by other cooperators, is still puzzling. Here, we investigate this question for a well-defined but generic situation based on typical life-cycles of microbial populations where individuals regularly form new colonies followed by growth phases.  We analyze two evolutionary mechanisms favoring cooperative behavior and study their strength depending on the inoculation size and the length of a life-cycle. In particular, we find that population bottlenecks followed by exponential growth phases strongly increase the survival and fixation probabilities of a single cooperator in a free-riding population. 
\end{abstract}


\section*{Introduction}
Cooperative behavior often provides a strong benefit for populations. But why are cooperators not undermined by non-cooperative individuals which take the benefit but save the costs for its provision~\cite{Hamilton:1964,Maynard,Okasha}? For higher developed organisms, there are several ways to escape this \emph{dilemma of cooperation}:  For individuals, which are capable to recognize other individuals, memorizing previous interactions and controlling their handling accordingly, reciprocity, and punishment can promote cooperation~\cite{Okasha,NowakCooperation}. However, those mechanisms cannot act in organisms of modest complexity like microbes or during the early course of life where memory and recognition were mostly lacking. 

In such scenarios, cooperation might prevail due to the structure of the population~\cite{Wright:1931,Hamilton:1964,Okasha,Fletcher:2009,Traulsen:2009,Julou:2013}; this idea has been studied both theoretically and experimentally in the context of kin- , group- and multilevel selection~\cite{Hamilton:1963,Hamilton:1964,MaynardSmith:1964,DSwilsongroup,Wade:1977,Wade:1978,Craig:1982,WILSON:1983, Goodnight:1985,Sober:1999,Fletcher:2004,Killingback:2006,Traulsen:2006a,Lehmann:2007,Fletcher:2007,West:2007a,Fletcher:2009,Hauert:2012}.  If cooperators  more likely interact with other cooperators (positive assortment), they keep most of their benefit for themselves and are less exploited by non-cooperators. However, due to the costs of cooperation, a fitness disadvantage compared to non-cooperators is still present: Positive assortment supports cooperation but is not necessarily sufficient to ensure its maintenance. Crucially, positive assortment can only act if cooperation is already established in the population such that cooperative individuals can successfully assort. Thus, the question remains how cooperation can emerge \emph{starting with a single cooperating mutant}.

In this manuscript, we address  this issue for a generic situation of microbial populations. Cooperative microbes typically produce public goods whose synthesis is metabolically costly~\cite{Velicer:2003p377,Kreft:2005,Brockhurst:2007,Gardner,Gore:2009,Hallatschek:2011,Buckling:2007}. For example, consider the proteobacteria \emph{Pseudomonas aeruginosa} and siderophore production: when iron is lacking in the environment, cooperative strains produce iron-scavenging molecules (siderophores)~\cite{Diggle, Buckling:2007}. Released into the environment these molecules can efficiently bind single iron molecules and the resulting complex can then be taken up by surrounding bacteria. Microbial populations are highly structured: several colonies form one population~\cite{Hall-Stoodley:2004,West:2006}. New colonies arise due to migration into new habitats or more actively due to controlled life-cycles triggered by environmental factors. For example, studies of \emph{P. aeruginosa}~\cite{Stoodley:2002,Hall-Stoodley:2004} confirm that  typical life-cycles pass through different steps with regularly occurring dispersal events ensued by the formation of new colonies.  As the initial colony sizes are typically small, such dispersal events coincide with population bottlenecks. 
Emulating the dynamical colony formation, microbial cooperation has been studied experimentally by employing a life-cycle where new colonies are regularly formed from old ones~\cite{Griffin,Chuang:2009}. These experiments and theoretical work~\cite{Hamilton:1964,MaynardSmith:1964,Okasha,Chuang:2010,Cremer:2011,Garcia2013} show that such a restructuring mechanism can cause an increase in the level of cooperation. 

Based on these observations, we here theoretically investigate if such a restructuring scenario cannot only maintain cooperation, but also allows for the evolution of cooperation from a single mutant. Assuming a constant population size, the onset of traits from a single  mutant and its fixation have been studied in the frameworks of population genetics and evolutionary game theory ~\cite{Kimura,Nowak:2004,Taylor,Traulsen:2005,Traulsen:2006,imhof,Antal,Cremer:2009,Parson:2010}. However the consequences of ecological factors such as population growth and population restructuring remain unclear~\cite{Fletcher:2004,Killingback:2006}.
To tackle this question for the microbial scenario introduced before, we consider a life-cycle model, consisting of three steps~\cite{MaynardSmith:1964,Cremer:2011}, see  Fig.~\ref{fig:cartoon}: 
\begin{figure}[!ht]
\centering
\includegraphics[width=.8\textwidth]{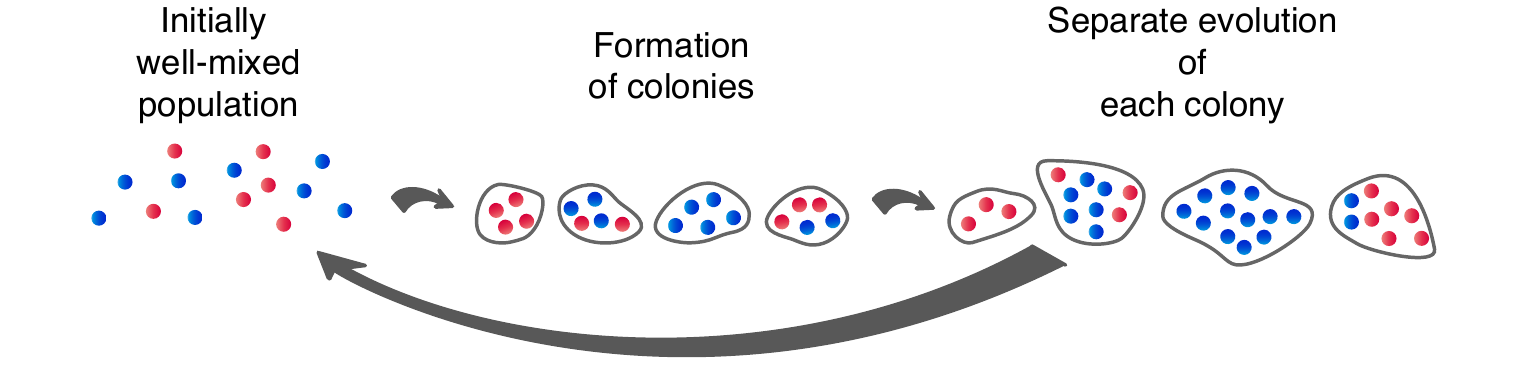}
\caption{{\bf A simplified life-cycle of microbial populations.} Cooperators  [blue] and free-riders [red] in an initially well-mixed population are randomly subdivided into groups of average size $n_0$. The groups then evolve separately following two main rules. First, groups with a higher fraction of cooperators grow faster as more public good is present in those populations. Second, in each group cooperators reproduce slower as free-riders as they do not have to provide the public good; for detail see main text. After a certain time $T$ all groups are merged and the cycle restarts.
}
\label{fig:cartoon}
\end{figure}
(i) In the \emph{group-formation step},  a population consisting of cooperators ($C$) and non-cooperating \emph{free-riders} ($F$) is randomly assorted into different groups (colonies). (ii) In the ensuing \emph{group-evolution step}, each group  evolves separately according to generic growth laws of microbes. (iii) In the \emph{group-merging step}, colonies are merged to one population again. Consecutively, the cycle starts anew with the next group-formation step. The synchronous reformation of groups via  merging all individuals into a single population follows recent experiments~\cite{Chuang:2009} and is obviously a simplification of natural colony formation. However, it captures the essence of regular occurring bottlenecks, namely rearranging colonies which have an initially small population size. Furthermore, as it comprises a worst-case scenario for cooperation, it is suited to study the possible onset of cooperation in microbial populations  starting with a single cooperative mutant. Starting with a well-mixed population with a fraction $x_0$ of cooperators, $M$ groups (colonies) are formed during the group-formation step. Successively, each group $i$ is assigned $\nu_{0,i}$ randomly chosen individuals of the well-mixed population; group sizes $\lbrace \nu_{0,i} \rbrace$ are Poisson-distributed with mean $n_0$. The random assortment leads to a statistical variation in the initial fraction of cooperators, $\xi_{0,i}$. Approximately, it is of the order $n_0\cdot x_0(1-x_0)$. We investigate this unbiased random assortment of groups as it does not assume individuals to be distinguishable by the sorting mechanism and again constitutes a worst-case scenario for cooperators.

After assortment into groups, each group $i$  evolves and grows separately. The dynamics within groups is given by a stochastic process based on birth and death events which are characterized by the corresponding per capita birth rates $\Gamma^+_{S,i}$ and death rates $\Gamma^-_{S,i}$, where $S\in\{C,F\}$ denotes the trait of the individual in group $i$~\cite{Melbinger:2010, Cremer:2011a}. The birth rates of individuals depend on two factors, namely the trait of the individual and the composition of the colony the individual is living in. First, cooperators reproduce slower than free-riders in each colony as they have metabolic costs due to the production of the public good, $\Gamma^+_{C,i}<\Gamma^+_{F,i}$. Second, as more cooperative groups produce more public good, individuals in colonies containing a higher fraction of cooperators are better off, $\Gamma^+_{S,i}<\Gamma^+_{S,j}$ for $\xi_i<\xi_j$. The death rates  incorporate the effect of limited resources and, therefore, increase with an increasing population size,  $\Gamma^-_{S,i}<\Gamma^-_{S,j}$ for $\nu_i<\nu_j$. For specificity, we assume the following birth and death rates which fulfill all conditions stated above:

\begin{align}
\Gamma^+_{S,i}=&r(1+p\,\xi_i)(1-\delta_{S,C}\,c)~\text{and }\Gamma^-_{S,i}=\nu_i/K \, ,
\label{eq:rates}
\end{align}
where $\delta_{S,C}$ is the Kronecker delta defined by $\delta_{C,C}=1$ and $\delta_{F,C}=0$.
While $p$ sets the growth advantage of cooperators on the colony level, the parameter $c$ measures the metabolic costs of cooperation. The growth rate $r\equiv 1$ is assumed to be fixed setting the time-scale of growth. The here assumed functional form of the growth rates reproduces the generically observed growth dynamics of microbial populations~\cite{Monod:1949}: Small colonies grow exponentially and their size is bounded by a maximal colony size which here scales with $K$. A more detailed description of the dynamics including a discussion of the deterministic equations can be found in Refs.~\cite{Melbinger:2010,Cremer:2011}. Furthermore, in~\cite{Cremer:2011} the specific form of the rates~\eqref{eq:rates} is  justified by successful comparisons with experiments by Chuang et al.~\cite{Chuang:2009}. Note also that the qualitative results presented in this manuscript do not depend on the specific functional forms of the growth rates, but only on the rather generic conditions of population bottlenecks followed by growth. 

After a \emph{regrouping time} $T$ the separated groups are merged again into one well-mixed population with a then changed global population size $N=\sum \nu_i$, and a fraction of cooperators which is given by the weighted average,
\begin{equation}
x = \sum_i \xi_i \nu_i / \sum_i \nu_i \, .
\label{eq:average}
\end{equation}
The cycle then starts anew with the new fraction of cooperators, $x_0\equiv x$. Although the fraction of cooperators within each group is expected to decrease during group-evolution, an increase in the global fraction of cooperators is possible in principle: The disadvantage of cooperation within each group can be overcome by changing weights, $\nu_i/N$, in the total population. To achieve this there must be a sufficiently high positive correlation between group size and cooperator fraction~\cite{Price:1970,Okasha}. Such an increase of cooperation is an example of Simpson's paradox~\cite{Okasha,Chuang:2009}. 

For the random assortment of groups considered here, two mechanisms promoting cooperation can be distinguished as previously studied~\cite{Cremer:2011}.  First, for very small population bottlenecks, purely cooperative colonies might be formed where there is no conflict with free-riders  (\emph{group-fixation mechanism}). Second, more cooperative groups grow comparably fast and thereby compensate for the selection advantage of free-riders (\emph{group-growth mechanism}). As those mechanisms are crucial for the understanding of our results concerning single mutants spreading in the population, we first repeat some arguments from Ref.~\cite{Cremer:2011} and additionally introduce analytic calculations  and a study of the key parameters  to support them. The second part of this article is devoted to the main question of the paper namely whether a single cooperative mutant which cannot rely on benefits provided by other cooperators has the chance to spread in the population.
\section*{Results}
\begin{figure}[!ht]
\centering
\includegraphics[width=.8\textwidth]{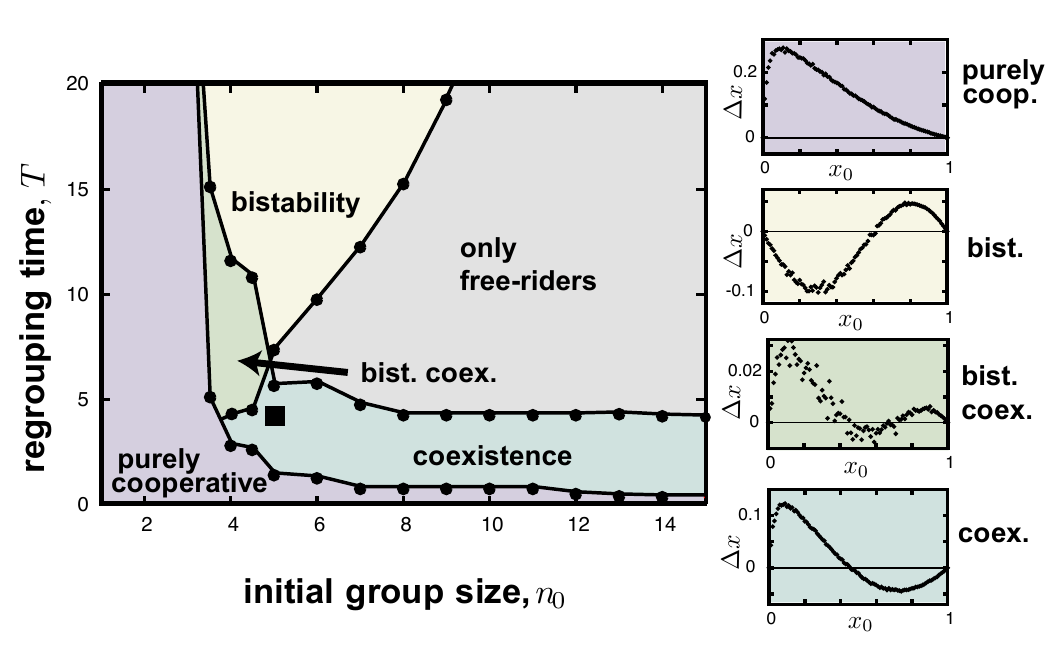}
\caption{{\bf Parameter dependence of the different cooperation scenarios.}
Depending on  $n_0$ and $T$ five different scenarios with different fixed point behaviors arise. For each scenario, we show an exemplary drift diagram on the right, where the change in the fraction of cooperators after one life-cycle is shown depending on the initial fraction of cooperators, $\Delta x( x_0)$. Dots correspond to simulation results of the transition points and the lines are guides to the eye to separate the different scenarios: Pure cooperation with a stable fixed point at 1, \emph{i.e.} for a purely cooperative population ($n_0=4$, $T=1.5$), coexistence with a stable fixed point at $0<x^*<1$ ($n_0=6$, $T=1.8$), bistability with an unstable fixed point at $0<x^*<1$  ($n_0=5$, $T=20$), bistable coexistence where both an stable and an unstable fixed point are present ($n_0=4$, $T=5.5$) and only free-riders. The black square corresponds to the parameters studied in Fig.~\ref{fig:traj} ( $n_0=5,~ T=4$). Other parameters are $p=10$, $K=100$, and $c=0.1$.
\label{fig:phases}
}
\end{figure}

In the following we analyze both mechanisms in detail, starting with the group-fixation mechanism. For long separate evolution of groups, $T\gg 1$, all groups reach a stationary state: They consist of either cooperators or free-riders only with a group sizes of $(1+p)(1-c)K$ and $K$, respectively. The global fraction of cooperators is then

\begin{align}
x'=\frac{(1+p)(1-c)P_C}{(1+p)(1-c)P_C+(1-P_C)} \, .
\label{eq:xprime}
\end{align}
$P_C$ denotes the probability for a group to consist of only cooperators after assortment.  In first order, only initially purely cooperative groups contribute to $P_C$ while all mixed groups are taken over by free-riders, such that  
\begin{equation}
 P_C=\frac{1}{e^{n_0}-1}\sum_{\nu_i=1}^{\infty}\frac{n_0^{\nu_i}}{\nu_i!}  x_0^{\nu_i}+\mathcal{O}(\frac{1}{K})=\frac{e^{n_0x_0}-1}{e^{n_0}-1}+\mathcal{O}(\frac{1}{K})\nonumber \, .
\end{equation}
If $x'$ exceeds the initial fraction of cooperators, $x_0$, the group-fixation mechanism is strong enough to overcome the advantage of free-riders. As $P_C$ increases with the initial fraction of cooperators, there is an unstable fixed point $x_u^*$, implicitly defined by $x_u^*=x'=x_0$ in Eq.~\eqref{eq:xprime}: For initial fractions, $x_0$, above $x_u^*$, a purely cooperative population is reached after several regrouping events. In contrast, when starting below, $x_0<x_u^*$, cooperators become extinct in the population.
This bistable behavior is illustrated in Fig.~\ref{fig:phases} where depending on the initial value $x_0$ the global fraction of cooperators is shown after a large regrouping time $T=20$. 

In contrast to the group-fixation mechanism, the group-growth mechanism acts for small times, where groups strongly grow. As cooperation enhances the growth speed of colonies, more cooperative groups have a larger weight in the average~\eqref{eq:average} even though $\dot \xi_i \leq 0$ holds in each group. Depending on the parameters, this positive effect is able to compensate for the selection disadvantage of cooperators. This can be quantified performing a van Kampen expansion of the master equation, see supplementary information. For binomial distributed groups, the change in the fraction of cooperators at time $t=0$ is given by,
\begin{equation}
\frac{d}{dt} x \propto\left[-c(1+px)+p/n_0\right]x(1-x) \, .
\end{equation}
The first term accounts for the selection advantage of free-riders for the growth advantage of more cooperative groups while the second one reflects the growth advantage of more cooperative groups.
The initial change, $\frac{d}{dt} x$, is larger if $x_0$ is small meaning that higher selection disadvantages can be overcome. Under regrouping after time $T$, the group-growth mechanisms results in a stable fixed point $x^*_S$. As the group-growth mechanisms relies only on variance in group composition, but not on the existence of purely cooperative groups, it acts for much stronger population bottlenecks, $n_0$, than the group-fixation mechanism does. However, as it is caused by population growth it can only act  for short regrouping times $T$. 

Repeated regrouping corresponds to an iterative map. The underlying dynamics (group formation and group evolution) result in
an effective drift: $\Delta x =x(T)-x_0$. Depending on the strength of both mechanisms, and thus $n_0$ and $T$, five distinct fixed point scenarios can be distinguished; examples for the corresponding stability plots ($\Delta x$) are shown in Fig.~\ref{fig:phases}. Besides the discussed \emph{bistable} [only group-fixation]  and \emph{stable coexistence scenario} [only  group-growth], there can be a \emph{bistable coexistence scenario} [group-growth and fixation mechanism]. In addition, there are the \emph{scenarios of only cooperation} and \emph{only defection}, where the sole stable fixed points are $x^*=1$ or $x^*=0$, respectively. 

\begin{figure}[!t]
\centering
\includegraphics[width=.8\textwidth]{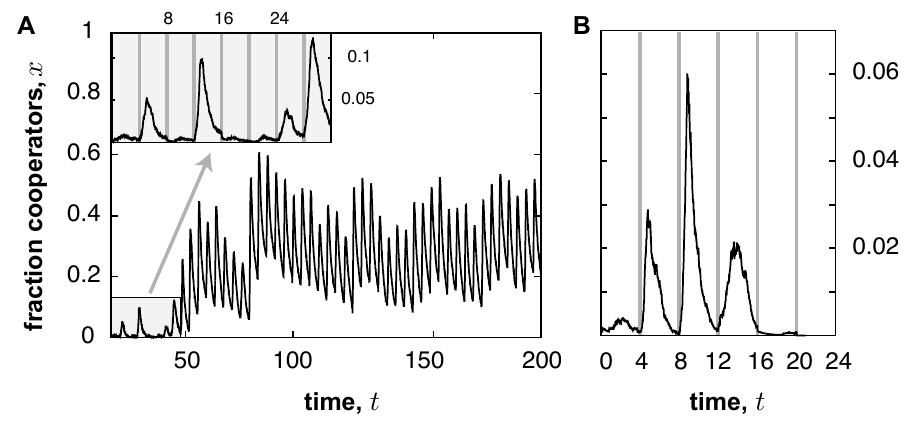}
\caption{{\bf Two exemplary realizations for the evolution of cooperation from a single mutant.} The parameters,  $T=4$ and $n_0=5$, correspond to a point in the coexistence regime indicated by the black square in Fig.~\ref{fig:phases}. In panel {\bf A} a coexistence trajectory is shown, while panel {\bf B} shows extinction of cooperators. For small times (zoom in inset  panel {\bf A}, and panel {\bf B}) cooperators have to survive a highly stochastic process:  Cooperators start with only one mutant, the population is subject to exponential growth which amplifies fluctuations,  and regrouping events add an additional level of stochasticity.  In addition, also for larger times the trajectory oscillates around the fixed point (for a detailed explanation see main text) and is therefore prone to extinction.
\label{fig:traj}
}
\end{figure}

To this point we have seen that the internal dynamics on the intra-group level result in an effective drift for the iterative map, which  can support cooperation.  We now want to examine whether this drift allows for the evolution of cooperation. In other words, we investigate if a single cooperative mutant can survive and spread in a free-riding population.
In Fig.~\ref{fig:traj} we show two exemplary trajectories for the time evolution of the fraction of cooperators. The parameters $T=4$ and $n_0=5$ corresponds to the coexistence regime in Fig.~\ref{fig:phases} (black square). The simulations are performed by placing a single cooperative mutant in one of the $M=200$ groups; this corresponds to an average initial fraction of cooperators of $x_0=1/(n_0M)=0.001$. In { Fig.~\ref{fig:traj}~\bf A}, a coexistence trajectory is shown while in Fig.~\ref{fig:traj} {\bf B}  cooperators go extinct at the fifth regrouping event ($t=20$). There are strong differences  between these trajectories and the ones known from non-linear dynamics models without an iterative map which are commonly used to analyze coexistence scenarios: Due to the dynamics during group-evolution, the fraction of cooperators does not only fluctuate slightly at the fixed point, but  oscillates around it. The reason is the interplay of the group-growth mechanism with the ensuing decline in the fraction of cooperators due to the selection disadvantage of cooperators in each group: After each regrouping event, the group growth mechanism causes an increase in the level of cooperators. Later on, when population growth slows down as groups  are already close to their carrying capacity, the group-growth mechanism is not effective anymore. Thus, the fraction of cooperators declines due to their growth disadvantage within each group. This dynamic pattern is then repeated in each life-cycle causing the oscillatory behavior. But not only oscillations increase the variability, also demographic fluctuations are particularly strong.  The reason is that small variations in the fraction of cooperators are exponentially amplified during group growth. Therefore, for large times ($t>100$ in  Fig.~\ref{fig:traj} {\bf A}) when the steady state is already reached the exact position of the maxima and minima in each life-cycle substantially varies between different regrouping events. In addition, also the regrouping mechanism  increases the randomness, \emph{i.e} due to the stochastic reformation of groups the fraction of cooperators can change during this process. In the inset of Fig.~\ref{fig:traj} {\bf A} the first eight regrouping steps are magnified. For instance during the second ($t=8$) and fourth ($t=16$) regrouping events  a substantial drop in the fraction of cooperators is present, while for other regrouping events it is hardly changed.

This analysis already shows, that the question whether a single mutant can survive is much less trivial than suggested by the phase diagram [Fig.~\ref{fig:phases}]: Mutants have to overcome two hampering factors which diminish the positive drift of the iterative map.  First, in each group free-riders are favored over cooperators due to the costs to produce the public goods $c$: During group-evolution,  the fraction of cooperators declines in each initially mixed group. Thus if group-evolution was not interrupted by the formation of new groups, cooperators would die out in those mixed groups. Therefore, in all parameter regimes (including the ones with a stable cooperation fixed point) the fixation probability of a free-rider is higher than the one of a cooperator during group-evolution. Second, even if cooperators have survived group-evolution, it is not assured that they survive regrouping as this process follows a random distribution. Especially for small fractions of cooperators, cooperators die out more likely than free-riders during regrouping. Taken together, it is not obvious whether the advantages of cooperators are sufficient to enable the evolution of cooperation from a single mutant.

\begin{figure}[!t]
\centering
\includegraphics[width=.8\columnwidth]{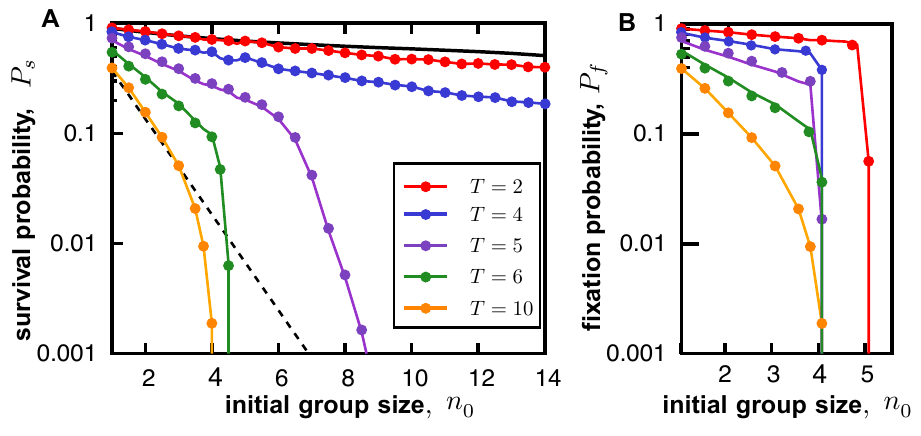}
\caption{
{\bf  Survival (panel A) and fixation probability (panel B) of cooperation.} When starting with a single mutant, the survival probability of cooperators, $P_s$, and their fixation probability, $P_f$, are shown after $500$ regrouping events.  Interestingly both probabilities are fairly large over a large parameter regime. Different colors correspond to different regrouping times: $T=\lbrace
2~(\textcolor{red}{\scriptsize  \bullet}),
4~(\textcolor{blue}{\scriptsize   \bullet}),
5~(\textcolor{Violet}{\scriptsize  \bullet}),
6~(\textcolor{darkgreen}{\scriptsize  \bullet}),
10~(\textcolor{BurntOrange}{\scriptsize \bullet})
\rbrace$, $M=200$, number of realizations $R\geq 10^5$. Other parameters: $p=10$, $K=100$, $c=0.1$. Black lines are calculated with our analytic approximation, Eq.~\eqref{eq:approx} for $T=2$ (solid) and $T\rightarrow\infty$ (dashed).}
\label{fig:evol_prob}
\end{figure}

In the following we study this issue by analyzing the \emph{fixation probability} $P_f$, and the \emph{survival probability} $P_s$. $P_f\equiv P_{\text{fix},C}$ denotes the probability that a single cooperative mutant will take over the entire population, whereas $P_s\equiv 1-P_{\text{fix},F}$ is the probability that free-riders have not fixated yet in the  population. Both probabilities are time-dependent, but reach quasi-stationary values under repeated regrouping, see supplementary information. Again starting with one cooperator, we performed  stochastic simulations of the population dynamics, each realization with $200$ regrouping events. In Fig.~\ref{fig:evol_prob}, the probabilities $P_s$ (panel {\bf A}) and $P_f$ (panel {\bf B}) depending on $n_0$ are shown for different regrouping times $T$. Due to the life-cycle dynamics the fixation behavior here is more intricate than for standard evolutionary dynamics. In those standard models, it is the population size only that determines the weight of  fluctuations with respect to the deterministic drift~\cite{Kimura}. In the limit of large population sizes fluctuations become irrelevant and a stable fixed point is reached with absolute certainty. In contrast, for our life-cycle model, the stability diagram and the deterministic drift themselves depend on the initial population size $n_0$: cooperators are less favored for higher values of $n_0$.
For small $n_0$, the regime of pure cooperation with a stable fixed point at $x^*=1$ is present, cf.~Fig.~\ref{fig:phases}. In accordance with that both fixation and survival probability are equal and fairly large. With increasing initial population sizes $n_0$ the strength of the group-growth mechanism is reduced and we observe a decline in the survival and fixation probabilities. This decline is faster for larger regrouping times since cooperators stay longer in the evolution step and the thereby created selection disadvantage, $s$, accumulates during the time interval $[0,T]$. Further increasing initial groups sizes cause a steep drop in the fixation probability. The reason is that upon increasing $n_0$ one is leaving the regime of pure cooperation, cf. Fig.~\ref{fig:phases} red line. Hence, the fixation probability of cooperators dramatically declines. While the so far discussed fixation probabilities only depend quantitatively on the regrouping time $T$, the survival probabilities are also qualitatively influenced by this parameter. For small $T$, the system is in the coexistence regime which implies that cooperators neither fully take over the population nor die out easily. Thus the survival probability shows only a slow decay with $n_0$ in this regime. In contrast, for larger regrouping times, the dynamics becomes unstable. Because the dynamics start with a single cooperative mutant, the group-fixation mechanism is not efficient as it is unlikely that the threshold above which cooperators are favored is reached. Therefore, the survival probability declines rapidly with increasing $n_0$.

Let us finally discuss how the survival probabilities can be understood, based on analytic arguments. To this end, we approximate the dynamics by considering the first life-cycle only, \emph{i.e.} the first group-evolution steps and the ensuing formation of new colonies. As most extinction events of cooperators happen at the beginning, this approximation captures the extinction dynamics qualitatively correctly, see supplementary information. The success of a cooperator crucially depends on the size of the group it is living in. For larger groups it has to compete with more free-riders and its survival chances are diminished. As the group sizes are Poisson-distributed, the probability that a cooperator emerges in a group of size $k$ is given by $n_0^{k-1}/(k-1)!e^{-n_0}$. The probability that cooperators survive the first regrouping step is the probability that not all $Mn_0$ newly formed groups are purely free-riding. For a realization with a fraction of $x$ cooperators before regrouping, this probability is given by $1-e^{-xn_0 M}$. Taken together this leads to the overall survival probability

\begin{align}
 P_S=e^{-n_0}\!\sum_{k=0}^\infty\! n_0^{k\!-\!1}\!/(k\!-\!1)! \langle1\!-\!e^{-x(k,\!T)n_0\!M}\rangle 
 \label{eq:approx}
\end{align}
where $\langle ..\rangle$ denotes  an average over all possible realizations. For large regrouping times, cooperators die out in all initially mixed groups, \emph{i.e.} $x=0$ for $k>1$. Thus, the survival probability simplifies to $e^{-n_0}$, which is in perfect agreement with our simulation results; see Fig~\ref{fig:evol_prob} (dashed black line). For smaller $T$, the group-growth mechanism is  dominant, and cooperators also survive in initially mixed groups leading to even higher values of the survival probabilities. For $T=2$, where the group-growth mechanism results in a stable fixed point, we compare  the approximation with the simulated survival probabilities and find good agreement; cf. Fig~\ref{fig:evol_prob} (solid black line).
\begin{figure}[!t]
\centering
\includegraphics[width=1\columnwidth]{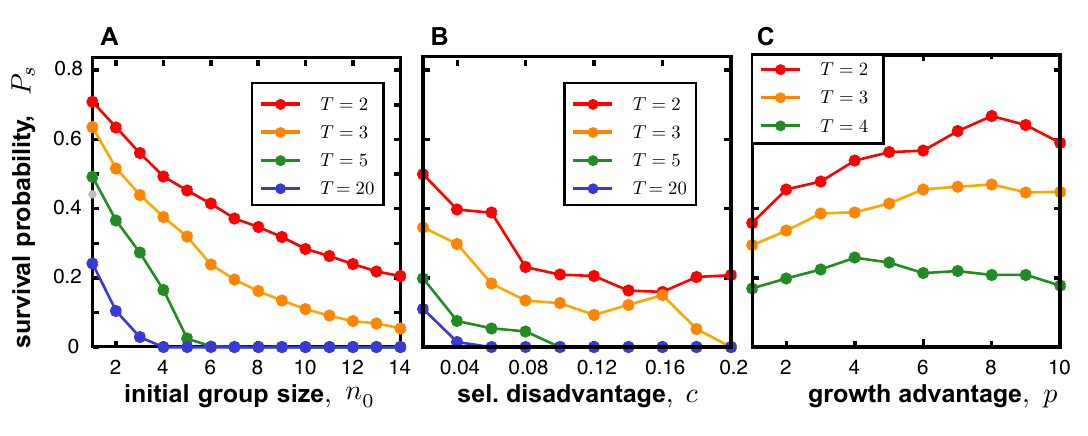}
\caption{ {\bf Robustness of the survival probabilities for different model parameters.} Panel {\bf A} shows the impact of exponentially distributed regrouping times depending on $n_0$. In panel {\bf B} the impact of varying selection disadvantages is investigated for $p=10$ and $n_0=5$.
In panel {\bf C} the survival probability for different values of $p$ scaling the growth advantage mediated by the public good is shown. Even for small values of $p$ the group-growth mechanism is effective and support the onset of cooperation. If not varied,  $n_0=5$ and $c=0.1$.  Other parameters are as in Fig.~4.
\label{fig:robust}}
\end{figure}

To further support the idea that ecological factors can explain the onset of cooperation, we study the robustness of our findings against changes of the model parameters. In particular we vary $p$ scaling the benefit of the public good, and $c$ measuring the metabolic costs due to the production of a public good. In Fig.~\ref{fig:robust}, we show the survival probabilities for different regrouping times  and $n_0=5$ depending on $c$ (panel {\bf B}) and $p$ (panel {\bf C}). Interestingly, already for comparably small values of $p$ the positive impact of a public good on the population size notably supports cooperation and thereby increases the survival probability. In contrast increasing costs $c$ decrease the survival chances, but the benefits of cooperation are strong enough to compensate for selection disadvantages of up to 20$\%$. In addition, we also demonstrate that our results do not depend on the assumption that the regrouping time is fixed. In Fig.~\ref{fig:robust} {\bf A}, survival probabilities are shown for regrouping times exponentially distributed with $\bar T$ according to $p(T)=\frac{1}{\bar T}\exp[-T/\bar T]$. Again, the survival probabilities are fairly large and our conclusion that a simple life-cycle supports cooperators strong enough to overcome their selection disadvantages is not changed. All in all, we therefore conclude  that our results are robust against changes in the details of the parameters and model assumptions and thereby constitute a possible mechanisms enabling the onset of cooperation.

\section*{Conclusion}
In this article, we studied the impact of ecological factors like population growth and population bottlenecks on the evolutionary dynamics of cooperating individuals. Our main findings can be subdivided into two blocks: First we analyzed the evolutionary dynamics acting on cooperators already abundant in a population, second we studied the survival chances of single mutants emerging in a purely non-cooperating environment. 

 In the first part, we were employing our recently introduced model~\cite{Cremer:2011} to study how a restructuring mechanism combined with typical growth conditions influences the evolutionary dynamics of public good producing bacteria. Depending on the inoculation size $n_0$ and the regrouping time $T$,  regimes of stable cooperation, coexistence between cooperators and free-riders and bistability emerge. Those regimes arise over a broad parameter regime even though the worst case scenarios for cooperators are assumed whenever model assumptions have to be made. Therefore, we believe that the mechanisms still apply in more realistic evolutionary scenarios where for example reassortment is not completely random or public goods are not equally distributed between all individuals~\cite{Julou:2013}. Other studies focussing on different aspects of the interplay between evolution and ecological factors support our findings: For instance, the impact of exponential growth following bottlenecks in infinite populations was shown to support cooperators~\cite{Fletcher:2004}. In Ref.~\cite{Killingback:2006} the competition between groups sharing a bounded global populations size and thereby competing for resources was investigated, while in Ref.~\cite{Brockhurst:2007} the impact of mutations on the beneficial effects of population growth for groups starting with only one individual was studied. In addition,  also the frequency of ecological disturbance and resource supply plays a crucial role for the resulting level of cooperation~\cite{Brockhurst:2007b,Brockhurst:2010}. Furthermore, beneficial effects for cooperators were also found when reassortment is not random but environmental driven~\cite{Pepper:2002,Pepper:2007,Ichinose2008221}. All studies emphasize the important role of ecological factors for understanding cooperation. In contrast to the mentioned studies, we focussed on a  description of bacterial growth, starting with an exponential growth phase reaching a carrying capacity later on. The different growth regimes (exponential growth and stationary state) influence the evolutionary dynamics differently: Both related mechanisms (group-growth and group-fixation) favor cooperators, but as confirmed by analytic arguments presented above, the strength of the group-growth mechanisms increases with small $x_0$ while the strength of group-fixation mechanism decreases. Therefore, both mechanisms can be associated with two different fixed point scenarios (stable and unstable). We present a full parameter study of the ensuing regimes of cooperation (fully cooperative, coexistence, bistability, bistable coexistence and purely free-riding)  for both key parameters, $n_0$ and $T$.
 
 In the second part of our paper we focussed on the question whether those beneficial mechanisms can also explain the onset of cooperation from a single mutant. Similar questions were extensively studied for evolutionary dynamics without regrouping where the factors influencing fixation probabilities for neutral, beneficial and deleterious mutations were investigated carefully, see e.g.~\cite{Ewens,KimuraOhta,Ohta2}. However, the non-iterative map caused by regrouping makes a new study essential as it alters many aspects of the evolutionary dynamics. Here, we demonstrate that ecological factors increase the survival and fixation probabilities of cooperators substantially. In particular, the group-growth mechanism allows for the  robust establishment of cooperation as it does not rely on a threshold fraction of cooperators to act effectively. Remarkably, the probability for a single mutant to succeed decreases only slowly with increasing $n_0$, and growth thus allows the onset of cooperation without the requirement of extremely narrow population bottlenecks.   Due to the robustness against parameter changes and the worst case assumptions employed, our model provides a proof of principle that ecological factors might explain the onset of cooperative behavior. Once cooperation is established in a population, more advanced mechanisms, which rely on cooperators already present in a population, like kin discrimination or other active forms of positive assortment, may evolve to further stabilize cooperative behavior, see \emph{e.g.}~\cite{Hamilton:1964, Queller:1992,Travisano:2004,Ostrowski:2008,Kuemmerli:2009, Velicer:2009,Julou:2013}.

\section*{Acknowledgments}
We thank Jan-Timm Kuhr  for discussion. Financial support by the Deutsche Forschungsgemeinschaft through the priority program ``Phenotypic heterogeneity and sociobiology of bacterial populations" (FR850/11-1) and the Nano Initiative Munich (NIM) is gratefully acknowledged.
calculations.

\begin{thebibliography}{10}

\bibitem{Hamilton:1964}
W.~D. Hamilton, ``The genetical evolution of social behaviour. {I+II},'' {\em
  J.Theor. Biol.}, vol.~7, p.~1, 1964.

\bibitem{Maynard}
J.~{Maynard Smith}, {\em Evolution and the Theory of Games}.
\newblock Cambridge: Cambridge University Press, 1982.

\bibitem{Okasha}
S.~Okasha, {\em Evolution and the Levels of Selection}.
\newblock Oxford: Oxford University Press, 2006.

\bibitem{NowakCooperation}
M.~A. Nowak, ``Five rules for the evolution of cooperation,'' {\em Science},
  vol.~314, p.~1560, 2006.

\bibitem{Wright:1931}
S.~Wright, ``Evolution in {M}endelian populations,'' {\em Genetics}, vol.~16,
  pp.~97--159, 1931.

\bibitem{Traulsen:2006a}
A.~Traulsen and M.~A. Nowak, ``Evolution of cooperation by multilevel
  selection,'' {\em Proc. Natl. Acad. Sci. U. S. A.}, vol.~103, no.~29,
  pp.~10952--10955, 2006.

\bibitem{Fletcher:2009}
J.~A. Fletcher and M.~Doebeli, ``A simple and general explanation for the
  evolution of altruism,'' {\em Proc R Soc Lond B}, vol.~276, pp.~13--19, 2009.

\bibitem{Traulsen:2009}
A.~Traulsen, ``Mathematics of kin- and group-selection: Formally equivalent?,''
  {\em Evolution}, vol.~64, pp.~316--323, 2009.

\bibitem{Julou:2013}
T.~Julou, T.~Mora, L.~Guillon, V.~Croquette, I.~J. Schalk, D.~Bensimon, and
  N.~Desprat, ``Cell-cell contacts confine public goods diffusion inside
  pseudomonas aeruginosa clonal microcolonies.,'' {\em Proceedings of the
  National Academy of Sciences}, 2013.

\bibitem{Hamilton:1963}
W.~D. Hamilton, ``The evolution of altruistic behavior,'' {\em Am. Nat.},
  vol.~97, pp.~354--56, 1963.

\bibitem{MaynardSmith:1964}
J.~{Maynard Smith}, ``Group selection and kin selection,'' {\em Nature},
  vol.~4924, p.~1145, 1964.

\bibitem{DSwilsongroup}
D.~S. Wilson, ``A theory of group selection,'' {\em Proc. Natl. Acad. Sci.
  USA}, vol.~72, no.~1, p.~143, 1975.

\bibitem{Wade:1977}
M.~J. Wade, ``An experimental study of group selection,'' {\em Evolution},
  vol.~31, no.~1, pp.~134--153, 1977.

\bibitem{Wade:1978}
M.~J. Wade, ``A critical review of the models of group selection,'' {\em The
  Quarterly Review of Biology}, vol.~53, no.~2, pp.~101--114, 1978.

\bibitem{Craig:1982}
M.~J. Wade, ``Group selection versus individual selection: an experimental
  analysis,'' {\em Evolution}, vol.~36, no.~2, pp.~271--282, 1982.

\bibitem{WILSON:1983}
D.~S. Wilson, ``The group selection controversy: History and current stauts,''
  {\em Ann Rev Ecol Syst}, vol.~14, pp.~159--186, 1983.

\bibitem{Goodnight:1985}
C.~J. Goodnight, ``The influence of environmental variation on group and
  individual selection in a cress,'' {\em Evolution}, vol.~39, no.~3,
  pp.~545--558, 1985.

\bibitem{Sober:1999}
E.~Sober and D.~S. Wilson, {\em Unto Others: The Evolution and Psychology of
  Unselfish Behavior}.
\newblock National Academy Press, Washington, 1999.

\bibitem{Fletcher:2004}
J.~Fletcher and M.~Zwick, ``Strong altruism can evolve in randomly formed
  groups,'' {\em J. Theor. Biol.}, vol.~228, pp.~303--313, Jan 2004.

\bibitem{Killingback:2006}
T.~Killingback, J.~Bieri, and T.~Flatt, ``Evolution in group-structured
  populations can resolve the tragedy of the commons,'' {\em Proc. R. Soc. B},
  vol.~273, p.~1477, 2006.

\bibitem{Lehmann:2007}
L.~Lehmann, L.~Keller, S.~West, and D.~Roze, ``Group selection and kin
  selection: Two concepts but one process,'' {\em Proc. Natl. Acad. Sci. U. S.
  A.}, vol.~104, no.~16, pp.~6736--6739, 2007.

\bibitem{Fletcher:2007}
J.~A. Fletcher and M.~Zwick, ``The evolution of altruism: game theory in
  multilevel selection and inclusive fitness,'' {\em J. Theor. Biol.},
  vol.~245, pp.~26--36, Mar 2007.

\bibitem{West:2007a}
S.~A. West, A.~S. Griffin, and A.~Gardner, ``Social semantics: how useful has
  group selection been?,'' {\em J. Evol. Biol.}, vol.~21, pp.~374--385, Nov
  2007.

\bibitem{Hauert:2012}
C.~Hauert and L.~Imhof, ``Evolutionary games in deme structured, finite
  populations,'' {\em J Theo Biol}, vol.~299, p.~106, 2011.

\bibitem{Velicer:2003p377}
G.~J. Velicer, ``Social strife in the microbial world,'' {\em Trends
  Microbiol.}, vol.~11, no.~7, p.~330, 2003.

\bibitem{Kreft:2005}
J.-U. Kreft and S.~Bonhoeffer, ``{The evolution of groups of cooperating
  bacteria and the growth rate versus yield trade-off},'' {\em Microbiol.},
  vol.~151, no.~3, pp.~637--641, 2005.

\bibitem{Brockhurst:2007}
M.~A. Brockhurst, ``Population bottlenecks promote cooperation in bacterial
  biofilms,'' {\em PLoS ONE}, vol.~2, p.~e634, Jan 2007.

\bibitem{Gardner}
A.~Gardner and K.~R. Foster, {\em The Evolution and Ecology of Cooperation -
  History and Concepts}.
\newblock Springer-Verlag, 2008.

\bibitem{Gore:2009}
J.~Gore, H.~Youk, and A.~van Oudenaarden, ``Snowdrift game dynamics and
  facultative cheating in yeast,'' {\em Nature}, vol.~459, p.~253, 2009.

\bibitem{Hallatschek:2011}
O.~Hallatschek, ``Noise driven evolutionary waves,'' {\em PLoS Comput Biol},
  vol.~7, p.~e1002005, 03 2011.

\bibitem{Buckling:2007}
A.~Buckling, F.~Harrison, M.~Vos, M.~A. Brockhurst, A.~Gardner, S.~A. West, and
  A.~Griffin, ``Siderophore-mediated cooperation and virulence in pseudomonas
  aeruginosa,'' {\em FEMS Microbiol. Ecol.}, vol.~62, no.~2, p.~135, 2007.

\bibitem{Diggle}
S.~P. Diggle, A.~S. Griffin, G.~S. Campbell, and S.~A. West, ``Cooperation and
  conflict in quorum-sensing bacterial populations,'' {\em Nature}, vol.~450,
  pp.~411--414, 2007.

\bibitem{Hall-Stoodley:2004}
L.~Hall-Stoodley, J.~W. Costerton, and P.~Stoodley, ``Bacterial biofilms: From
  the natural envrionment to infectious diseases,'' {\em Nat. Rev. Micro.},
  vol.~2, pp.~95--108, 2004.

\bibitem{West:2006}
S.~A. West, A.~S. Griffin, A.~Gardner, and S.~P. Diggle, ``Social evolution
  theory for microorganisms,'' {\em Nat. Rev. Microbiol.}, vol.~4, no.~8,
  p.~597, 2006.

\bibitem{Stoodley:2002}
P.~Stoodley, K.~Sauer, D.~G. Davies, and J.~W. Costerton, ``Biofilms as complex
  differentiated communities,'' {\em Ann. Rev. Microbiol.}, vol.~56,
  pp.~187--209, 2002.

\bibitem{Griffin}
A.~S. Griffin, S.~A. West, and A.~Buckling, ``Cooperation and competition in
  pathogenic bacteria,'' {\em Nature}, vol.~430, p.~1024, 2004.

\bibitem{Chuang:2009}
J.~S. Chuang, O.~Rivoire, and S.~Leibler, ``Simpson's paradox in a synthetic
  microbial system,'' {\em Science}, vol.~323, pp.~272--275, 2009.

\bibitem{Chuang:2010}
J.~S. Chuang, O.~Rivoire, and S.~Leibler, ``Cooperation and {H}amilton's rule
  in a simple synthetic microbial system,'' {\em Molec. Syst. Biol.}, vol.~6,
  p.~398, 2010.

\bibitem{Cremer:2011}
J.~Cremer, A.~Melbinger, and E.~Frey, ``Growth dynamics and the evolution of
  cooperation in microbial populations,'' {\em Scientific Reports}, vol.~2,
  p.~281, 2012.

\bibitem{Garcia2013}
T.~Garcia and S.~De~Monte, ``Group formation and the evolution of sociality,''
  {\em Evolution}, vol.~67, no.~1, pp.~131--141, 2013.

\bibitem{Kimura}
M.~Kimura, {\em The Neutral Theory of Molecular Evolution}.
\newblock Cambridge: Cambridge University Press, 1983.

\bibitem{Nowak:2004}
M.~A. Nowak, A.~Sasaki, C.~Taylor, and D.~Fudenberg, ``Emergence of cooperation
  and evolutionary stability in finite populations,'' {\em Nature}, vol.~428,
  pp.~646--650, 2004.

\bibitem{Taylor}
C.~Taylor, D.~Fudenberg, A.~Sasaki, and M.~A. Nowak, ``Evolutionary game
  dynamics in finite populations,'' {\em Bull. Math. Biol.}, vol.~66, no.~6,
  pp.~1621--1644, 2004.

\bibitem{Traulsen:2005}
A.~Traulsen, J.~C. Claussen, and C.~Hauert, ``Coevolutionary dynamics: from
  finite to infinite populations,'' {\em Phys. Rev. Lett.}, vol.~95, p.~238701,
  2005.

\bibitem{Traulsen:2006}
A.~Traulsen, M.~A. Nowak, and J.~M. Pacheco, ``Stochastic dynamics of invasion
  and fixation,'' {\em Phys. Rev. E}, vol.~74, no.~1, p.~011909, 2006.

\bibitem{imhof}
L.~A. Imhof and M.~A. Nowak, ``Evolutionary game dynamics in a
  {W}right-{F}isher process,'' {\em Journal of Mathematical Biolology},
  vol.~52, pp.~667--681, 2006.

\bibitem{Antal}
T.~Antal and I.~Scheuring, ``Fixation of strategies for an evolutionary game in
  finite populations,'' {\em Bull. Math. Biol.}, vol.~68, p.~1923, 2006.

\bibitem{Cremer:2009}
J.~Cremer, T.~Reichenbach, and E.~Frey, ``The edge of neutral evolution in
  social dilemmas,'' {\em New J. Phys.}, vol.~11, p.~093029, 2009.

\bibitem{Parson:2010}
T.~L. Parson, C.~Quince, and J.~B. Plotkin, ``Some consequences of demographic
  stochasticity in population genetics,'' {\em Genetics}, vol.~185,
  pp.~1345--1354, 2010.

\bibitem{Melbinger:2010}
A.~Melbinger, J.~Cremer, and E.~Frey, ``Evolutionary dynamics in growing
  populations,'' {\em Phys. Rev. Lett.}, vol.~105, p.~178101, 2010.

\bibitem{Cremer:2011a}
J.~Cremer, A.~Melbinger, and E.~Frey, ``Evolutionary and population dynamics: A
  coupled approach,'' {\em Phys. Rev. E}, vol.~84, p.~051921, 2011.

\bibitem{Monod:1949}
J.~Monod, ``The growth of bacterial cultures,'' {\em Annu. Rev. Microbiol.},
  vol.~3, pp.~371--394, 1949.

\bibitem{Price:1970}
G.~R. Price, ``Selection and covariance,'' {\em Nature}, vol.~227, p.~520,
  1970.

\bibitem{Brockhurst:2007b}
M.~A. Brockhurst, A.~Buckling, and A.~Gardner, ``Cooperation peaks at
  intermediate disturbance,'' {\em Curr. Biol.}, vol.~17, no.~761, 2007.

\bibitem{Brockhurst:2010}
M.~A. Brockhurst, M.~G. J.~L. Habets, B.~Libberton, A.~Buckling, and
  A.~Gardner, ``Ecological drivers of the evolution of public-goods cooperation
  in bacteria,'' {\em Ecology}, vol.~91, no.~2, pp.~334--340, 2010.

\bibitem{Pepper:2002}
J.~W. Pepper and B.~Smuts, ``A mechanism for the evolution of altruism among
  nonkin: positive assortment through environmental feedback,'' {\em Am. Nat.},
  vol.~160, no.~2, pp.~205--13, 2002.

\bibitem{Pepper:2007}
J.~W. Pepper, ``Simple models of assortment through environmental feedback,''
  {\em Artif. Life}, vol.~13, no.~1, pp.~1--9, 2007.

\bibitem{Ichinose2008221}
G.~Ichinose and T.~Arita, ``The role of migration and founder effect for the
  evolution of cooperation in a multilevel selection context,'' {\em Ecological
  Modelling}, vol.~210, no.~3, pp.~221 -- 230, 2008.

\bibitem{Ewens}
W.~J. Ewens, {\em Mathematical Population Genetics}.
\newblock Springer, 2nd.~ed., 2004.

\bibitem{KimuraOhta}
M.~Kimura and T.~Ohta, ``The average number of generations until fixation of a
  mutant gene in a finite population,'' {\em Genetics}, vol.~61, pp.~763--771,
  1969.

\bibitem{Ohta2}
T.~Ohta, ``Slightly deleterious mutant substitutions in evolution,'' {\em
  Nature}, vol.~246, p.~96, 1973.

\bibitem{Queller:1992}
D.~Queller, ``Does population viscosity promote kin selection,'' {\em Trends
  Ecol Evol}, vol.~7, p.~322, 1992.

\bibitem{Travisano:2004}
M.~Travisano and G.~J. Velicer, ``Strategies of microbial cheater cotrol,''
  {\em Trends Microbiol.}, vol.~12, no.~2, pp.~72--77, 2004.

\bibitem{Ostrowski:2008}
E.~Ostrowski, M.~Katoh, G.~Shaulsky, D.~Queller, and J.~E. Strassmann, ``Kin
  discrimination increases with genetic distance in a social amoeba,'' {\em
  PLoS Biol.}, vol.~6, no.~11, p.~e287, 2008.

\bibitem{Kuemmerli:2009}
R.~K{\"u}mmerli, A.~S. Griffin, S.~A. West, A.~Buckling, and F.~Harrison,
  ``Viscous medium promotes cooperation in the pathogenic bacterium
  {P}seudomonas aeruginosa,'' {\em Proc Biol Sci}, vol.~276, p.~3531, 2009.

\bibitem{Velicer:2009}
G.~J. Velicer and M.~Vos, ``Sociobiology of the {M}yxobacteria,'' {\em Annu.
  Rev. Microbiol.}, vol.~63, pp.~599--623, 2009.

\end{thebibliography}

\begin{thebibliography}{1}
\bibitem{vanKampen:2001}
{Van Kampen} N (2001) Stochastic Processes in Physics and Chetry (North-Holland
  Personal Library).
\newblock North Holland, 2nd edition.

\bibitem{Kimura}
Kimura M (1983) The Neutral Theory of Molecular Evolution.
\newblock Cambridge: Cambridge University Press.

\end{thebibliography}

\newpage
\renewcommand{\theequation}{S\arabic{equation}}
\renewcommand{\thefigure}{S\arabic{figure}}
\setcounter{figure}{0} 
\setcounter{equation}{0} 

{\LARGE \bf  Supporting Information} \newline ~\newline
In this supplementary material, we show how van Kampen's system size expansion can be used to derive the initial slope of the fraction of cooperators, Eq.~(4) main text. In addition, we present data which confirms the validity of Eq.~(4) and  approximations made for the calculations shown in the main text.

\section{Van Kampen Expansion}
As already described in the main text, the group growth mechanism is a fluctuation driven effect. To describe it, we perform an Omega expansion~\cite{vanKampen:2001} of the master equation (ME) defined by the birth and death rates Eqs.~(1) (main text). The results obtained in this section hold for a fixed initial group size, where the random assortment only influences the composition but not the number of individual of a colony. In most of our simulations (all except Fig.~S1), we use Poisson distributed group sizes and fluctuations in the initial composition of a group are additionally enhanced by variation in group-size, see main text. This full solution can be calculated by weighting the result for fixed colony sizes by a Poisson distribution with mean, $n_0$. But as the important parameter dependences are already present when considering only binomial distributed groups and the expression are less lengthy, we focus on this simpler case in the following. Before we start with the calculations, we also want to make another comment. Normally, the condition for successfully performing the Omega expansion, is a stable deterministic attractor. However, here we do not want to analyze the resulting Fokker-Planck equation in detail, but only use it to determine an estimate of the strength of the group-growth mechanism at $t\rightarrow0$ (if the mechanism is not strong enough to overcome the selection pressure at the very beginning, it will also not happen later as fluctuations become smaller with increasing time). Therefore, the strict assumption of a stable fixed point  can be relaxed, which leads to good results as we demonstrate at the end of this section.
The ME describing the evolutionary dynamics is given by,
\begin{align}
\partial_t P(\!N_C\!,\!N_F\!)=&[(\mathbf{E}_C^--1)r(1+p\frac{N_C}{N_C+N_F})N_C
+(\mathbf{E}_F^--1)r(1+p\frac{N_C}{N_C+N_F})(1+s)N_F\nonumber\\
+&(\mathbf{E}_C^+\!-\!1)N_C\frac{N_C+N_F}{K}
+(\mathbf{E}_F^+\!-\!1)N_F\frac{N_C+N_F}{K}]P(\!N_C\!,\!N_F\!),
\label{eq:ME}
\end{align}
where $P(N_C,N_F)$ is the probability to find a group containing $N_C$ cooperators and $N_F$ free-riders. The ladder operators $\mathbf{E}^\pm_{S}$ are defined by $\mathbf{E}^\pm_{S}f(N_S)=f(N_S\pm1)$.

Now, within the van Kampen system size expansion, the stochastic variables, $N_C$ and $N_F$, are substituted by a linear combination of functions describing their rescaled means, $c(t)$ and $f(t)$,  and new stochastic variables, $\xi$ and $\mu$, accounting for  fluctuations,
\begin{align}
N_C=&Kc(t)+\sqrt{K}\xi,\nonumber\\
N_F=&Kf(t)+\sqrt{K}\mu.\label{eq:subs}
\end{align}
 In this expressions fluctuations are already weighted according the well-known square root dependence on the population size~\cite{Kimura} which scales with the carrying capacity $K$.
 
 Employing~Eqs.~\eqref{eq:subs}   in Eq.~\eqref{eq:ME} and expanding the resulting expression in the carrying capacity $K$ leads to the mean-field equations for $c(t)$ and $f(t)$ in order $\sqrt K$ and a Fokker-Planck equation (FPE) for $P(\xi,\mu)$ in higher orders. As the FPE is lengthy and does not contributed to the understanding of our calculation, we do no write it down explicitly here.  In Table~\ref{table}, we summarize all changes arising with the substitution and the expansions which are used to transform the ME to a FPE equation. 

  \renewcommand{\arraystretch}{1.5}

 \begin{table}[htdp]
\begin{center}
\begin{tabular}{|c|c|}
\hline
original form & substitution and expansion\\
\hline\hline
$\partial_t P(N_C,N_F) $ & $~\partial_t  P(\xi,\mu)-\sqrt{K}(\dot c(t)\partial_\xi \tilde P(\xi,\mu)+\dot f(t)\partial_\mu  P(\xi,\mu)~$\\ \hline
$\mathbf{E}^\pm_{S}$ & $~1\pm\frac{1}{\sqrt{K}}\partial_{\zeta}+\frac{1}{2K}\partial^2_\zeta~$ \\ \hline
$N_{S}$&  $Ks(t)+\sqrt{K}\zeta$ \\ \hline
$N_{S}^2$& $K^2\sigma(t)^2+2K^{3/2}\sigma(t)\zeta+K\zeta^2$\\\hline
$\frac{N_{C}}{N_C+N_F}$& $x(t)+\frac{1}{m(t)\sqrt{K}}\left[(1-x(t))\xi-x(t)\mu\right]$ \\\hline

\end{tabular}
\caption{\label{table}Table summarizing the substitutions and  expansions which are employed to transform the ME into a Fokker-Planck equation. The stochastic variable $\zeta$ represents the fluctuations $\zeta=\{\xi,\mu\}$. The means are either given by $\sigma(t)=\{c(t),f(t)\}$ or $x(t)=\frac{c(t)}{c(t)+f(t)}$ and $m(t)=c(t)+f(t)$.}
\end{center}
\end{table}%

\begin{figure}

\includegraphics[width=.5\textwidth]{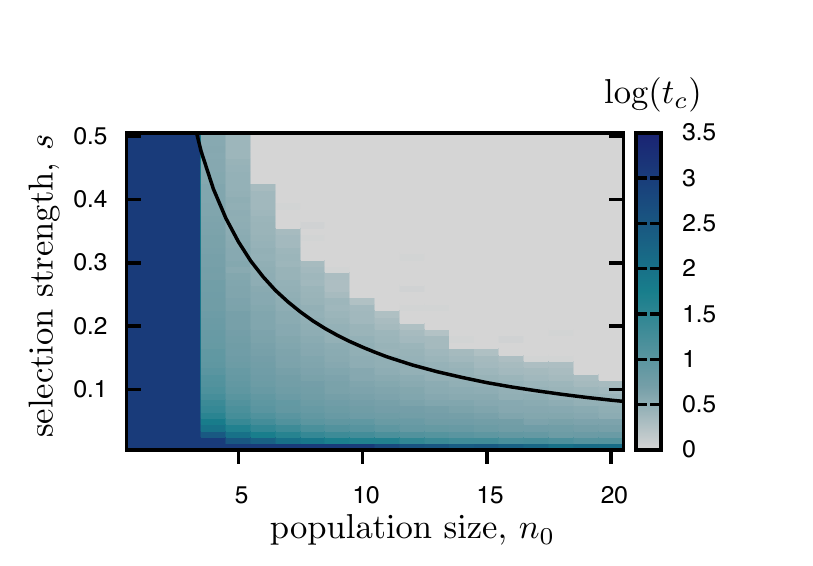}
\centering
\caption{Condition for the increase in the level of cooperation caused by the group-growth mechanism. The cooperation time is shown for different values for $s$ and $n_0$. The black line corresponds to the condition,~\eqref{eq:condition}. The other parameters are $p=10,~r=1$ and $K=100$.\label{fig:condition}}
\end{figure}

\begin{figure}

\includegraphics[width=.5\textwidth]{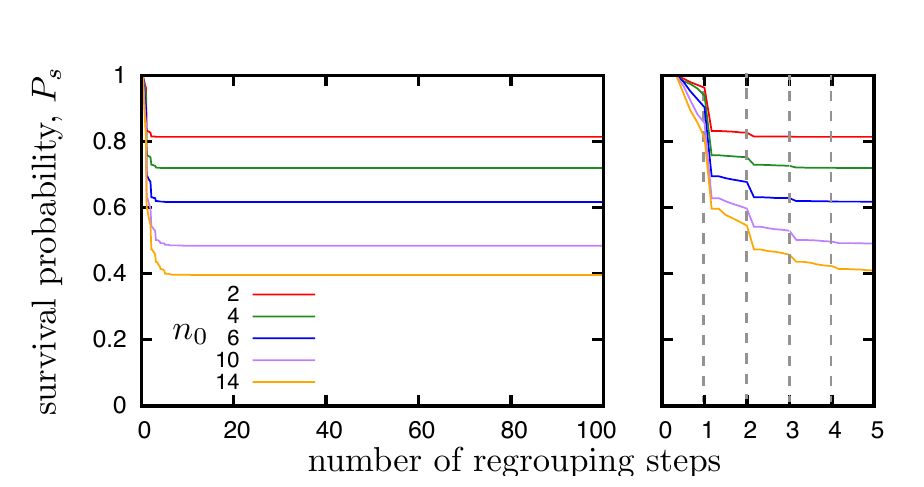}
\centering
\caption{Temporal change of the survival probability of cooperators, $P_S$, when starting with a single mutant. Change occurs only during the first regrouping events. Lines show $P_S$ for different initial group sizes $n_0$ and fixed regrouping time $T=2$.\label{fig:time_course}}
\end{figure}

 Next, we analyze the mean values and second moments of the fluctuations. They can be calculated by multiplying  the FPE with the respective fluctuation variable and integrating over both, $\int d\xi d\mu~ \zeta ...$ with $\zeta=\{\xi,\,u\}$. At time $t=0$, the means of the fluctuations vanish according to the initial conditions $\langle\xi_0\rangle=\langle\mu_0\rangle=0$.
 Thus the time evolution of the means is determined by only the second moments and given by,
  \begin{align}
 \langle \dot \xi \rangle =&+\!\frac{rp}{m_0\sqrt{K}}\left[(1\!-\!x_0)^2\langle\xi^2_0\rangle\!-\!2x_0(1\!-\!x_0)\langle\xi\mu_0\rangle\!+\!x_0^2\langle\mu_0^2\rangle\right], \nonumber \\
 \langle \dot \mu \rangle =&-\!\frac{rp(1+s)}{m_0\sqrt{K}}\left[(1\!-\!x_0)^2\langle\xi^2_0\rangle\!-\!2x_0(1\!-\!x_0)\langle\xi\mu_0\rangle\!+\!x_0^2\langle\mu_0^2\rangle\right], \nonumber\\
 \label{eq:first}
 \end{align}
  where we used $m_0=c(0)+f(0)$ and $x_0=c_0/m_0$. As the composition of each group is drawn from a binomial distribution, the variance of this distribution sets the initial condition for the second fluctuation moments, $\langle \xi_0^2\rangle=\langle \mu_0^2\rangle=-\langle\xi\mu_0\rangle=nx_0(1-x_0)$. Employing this in Eq.~\eqref{eq:first} leads to,
 
 \begin{align}
 \langle \dot \xi \rangle =&+\frac{rpx_0(1-x_0)}{\sqrt{K}},\nonumber \\
 \langle \dot \mu \rangle =&-\!\frac{rpx_0(1-x_0)(1\!+\!s)}{\sqrt{K}}.
 \label{eq:first_mod}
 \end{align}
The reason for the different signs of both differential equations is that the global growth rate $1+px$ is positively correlated with the fraction of cooperators $x$ while it is negatively correlated with the fraction of free-riders, $1-x$.  Hence, cooperators are favored by fluctuations. To compare the strength of this fluctuation driven mechanism with the deterministic  selection disadvantage, we analyze the average fraction of cooperators. To first order in $1/K$, it is given by,
\begin{align}
\langle x \rangle=&\left\langle \frac{x+\frac{1}{\sqrt{K}}\xi}{n+\frac{1}{\sqrt{K}}(\xi+\mu)}\right\rangle\nonumber \\ \approx &x(t)+\frac{1}{\sqrt{K}}\left((1-x)\langle\xi\rangle-x\langle\mu\rangle\right).
\end{align}
Differentiating this expression, neglecting higher order of $s$ and $1/K$, and using Eqs.~\eqref{eq:first_mod} leads to,

 \begin{align}
 \frac{d}{dt}\langle x \rangle=&-sr(1+px_0)x_0(1-x_0)+\frac{rpx_0(1-x_0)}{m_0K}=\nonumber \\ =&-sr(1+px_0)x_0(1-x_0)+\frac{rpx_0(1-x_0)}{n_0}.
 \end{align}
To confirm the validity of this calculation, we determine the parameter regime, in which the positive second term can overcome the selection disadvantage caused by cooperators $ \frac{d}{dt}\langle x \rangle>0$. This condition holds if,
\begin{align}
s>\frac{p}{n_0(1+px_0)}
\label{eq:condition}
\end{align}
 In Fig.~\ref{fig:condition}, we compare this condition to simulation data, for varying $s$ and $n_0$. We use the cooperation time, the time until the fraction of cooperators drops under its initial value, as a measure. If the cooperation time is zero, the advantage due to fluctuations is not large enough to overcome the selection disadvantage, while for finite values of the cooperation time, the group-growth mechanism is sufficient to compensate for slower growth of cooperators. As shown in Fig.~\ref{fig:condition},  Eq.~\eqref{eq:condition} is in agreement with the position where the cooperation time vanishes. Note that the analytical solution derived before gives a lower bound for the the fluctuation driven enhancement of cooperation: We only take into account the initial fluctuations given by a binomial distribution. In addition, demographic fluctuations further enhance cooperation, explaining the systematic shift in Fig.~S1.


\section{Time-Dependence Survival Probability}

In Fig.~3 in the main text, we show the extinction probabilities after 100 regrouping steps. Here, we want to confirm that after this time, already a quasi stationary state is reached. In Fig.~\ref{fig:time_course}, the time evolution of the survival probability of cooperators is shown for exemplary parameters. The data clearly shows that, as already mentioned in the body of this paper, the first regrouping steps mainly determine the fate of the population. If a single cooperative mutant survives those first grouping events, its extinction probability almost vanishes. Besides validating that after $100$ regrouping steps already a quasi steady state is reached, the presented data also justifies our analytical approach to estimate survival probabilities based only on the first regrouping event.

\end{document}